\documentclass{article}

    \PassOptionsToPackage{numbers, compress}{natbib}
   \usepackage[nonatbib,preprint]{neurips_2024}

\usepackage[utf8]{inputenc} % allow utf-8 input
\usepackage[T1]{fontenc}    % use 8-bit T1 fonts
\usepackage{hyperref}       % hyperlinks
\usepackage{url}            % simple URL typesetting
\usepackage{booktabs}       % professional-quality tables
\usepackage{amsfonts}       % blackboard math symbols
\usepackage{nicefrac}       % compact symbols for 1/2, etc.
\usepackage{microtype}      % microtypography
\usepackage{xcolor}         % colors
\usepackage{cite}
\usepackage{latexsym} 	
\usepackage{amssymb} 
\usepackage{amsbsy}
\usepackage{amsmath}
\usepackage{epsfig}     
\usepackage{graphics,color}
\usepackage{multirow}
\usepackage{subcaption}
\usepackage{changes}

\newcommand{\be}{\begin{equation}}
\newcommand{\ee}{\end{equation}}
\newcommand{\bea}{\begin{eqnarray}}
\newcommand{\eea}{\end{eqnarray}}
\newcommand{\bean}{\begin{eqnarray*}}
\newcommand{\eean}{\end{eqnarray*}}
\newcommand{\bit}{\begin{itemize}}
\newcommand{\eit}{\end{itemize}}

\hypersetup{linkcolor=blue}   %All internal links

\title{Diffusion models for lattice gauge field simulations}

\author{
Qianteng Zhu\\
INPAC, Key Laboratory for Particle Astrophysics and Cosmology (MOE),\\
Shanghai Key Laboratory for Particle Physics and Cosmology,\\
School of Physics and Astronomy, Shanghai Jiao Tong University, Shanghai 200240, China \\
\textit{zhuqianteng@sjtu.edu.cn}
\And
Gert Aarts \\
Department of Physics, Swansea University, Swansea, SA2 8PP,  United Kingdom \\
\textit{g.aarts@swansea.ac.uk}
\And
Wei Wang \\
INPAC, Key Laboratory for Particle Astrophysics and Cosmology (MOE),\\
Shanghai Key Laboratory for Particle Physics and Cosmology,\\
School of Physics and Astronomy, Shanghai Jiao Tong University, Shanghai 200240, China \\
Southern Center for Nuclear-Science Theory (SCNT),\\
Institute of Modern Physics, Chinese Academy of Sciences,\\
96 South Sihuan Rd. Huicheng District, Huizhou 516000, Guangdong, China\\
\textit{wei.wang@sjtu.edu.cn}
\And
Kai Zhou \\
School of Science and Engineering, \\
The Chinese University of Hong Kong, Shenzhen (CUHK-Shenzhen), Guangdong, 518172, China  \\
Frankfurt Institute for Advanced Studies, D-60438, Frankfurt am Main, Germany \\
\textit{zhoukai@cuhk.edu.cn}
\And
Lingxiao Wang \\
Interdisciplinary Theoretical and Mathematical Sciences Program (iTHEMS),\\
RIKEN, Wako, Saitama 351-0198, Japan \\
\textit{lingxiao.wang@riken.jp}
 }

\begin{document}

\maketitle

\begin{abstract}
We develop diffusion models for lattice gauge theories which build on the concept of stochastic quantization. This framework is applied to $U(1)$ gauge theory in $1+1$ dimensions. We show that a model trained at one small inverse coupling can be effectively transferred to larger inverse coupling without encountering issues related to topological freezing, i.e., the model can generate configurations corresponding to different couplings by introducing the Boltzmann factors as physics conditions, while maintaining the correct physical distributions without any additional training. This demonstrates the potential of physics-conditioned diffusion models for efficient and flexible lattice gauge theory simulations.

\end{abstract}

%%%%%%%%%%%%%%%%%%%%%%%%%%%%%%%%%%%%%%%%%%%%%%%%%%%

\section{Introduction}
Lattice Quantum Chromodynamics (QCD) is a crucial non-perturbative framework for studying strong interactions, offering valuable insight into hadronic structure and the quark-gluon plasma~\cite{Ding:2015ona,Gross:2022hyw,Aarts:2023vsf}. However, as Monte Carlo simulations approach the continuum limit, they are increasingly hindered by topological freezing, where large energy barriers between topological sectors degrade sampling efficiency and dramatically increase autocorrelation times~\cite{Alles:1996vn,DelDebbio:2004xh,Albandea:2021lvl}.
While several methods have been proposed to enhance Monte Carlo algorithms~\cite{Albandea:2021kwe,Albandea:2021lvl,Cossu:2021bgn}, the energy function-dependent nature of these approaches presents a fundamental limitation, preventing efficient sampling across different regions of parameter space.

To address these challenges, we propose a diffusion model-based generative algorithm for lattice gauge theory, which builds on the concept of stochastic quantization \cite{Parisi:1980ys,Damgaard:1987rr,Namiki:1993fd}. By training the model on gauge field configurations obtained at a single gauge coupling and volume, we demonstrate its ability to generalize for different parameters without the need for retraining, effectively bypassing issues related to topological freezing and scalability. This novel approach provides an interpretable and scalable solution for global sampling, offering a promising alternative to traditional Monte Carlo methods in lattice field theory.

\paragraph{Related work}  
Recent progress in generative models has introduced new possibilities for lattice field theory simulations~\cite{Cranmer:2023xbe}. Flow-based models, a prominent explicit likelihood estimation method, have gained attention for enabling global sampling in lattice simulations due to their invertibility and explicit use of gauge equivariance~\cite{Kanwar:2020xzo,Boyda:2020hsi,Kanwar:2021wzm,Albergo:2021vyo,Cranmer:2023xbe,Kanwar:2024ujc}. Besides, some variants of the normalising flow have also been developed recently, such as continuous normalizing flow~\cite{chen:2018,deHaan:2021erb,Gerdes:2022eve,Caselle:2023mvh} and stochastic normalizing flow \cite{wu2020stochastic,Caselle:2022acb}. Diffusion models have recently excelled in generating high-quality samples across diverse fields~\cite{yang:2022diffusion,croitoru2023diffusion}, including high-energy physics~\cite{Mikuni:2022xry,Amram:2023onf,Mikuni:2023dvk,Devlin:2023jzp}. The application to lattice field theory was initiated in Refs.~\cite{Wang:2023exq,Wang:2023sry}, where the connection to stochastic quantization~\cite{Parisi:1980ys,Damgaard:1987rr,Namiki:1993fd} was highlighted;  a Feynman path integral formulation was presented later~\cite{Hirono:2024zyg}.

\section{Physics-conditioned diffusion models}

Diffusion models can be realized \cite{Wang:2023exq} as a variant of stochastic quantization \cite{Parisi:1980ys}, an alternative method for quantizing field theories especially useful for gauge theories \cite{Damgaard:1987rr,Namiki:1993fd}.
One introduces a fictitious time $\tau$ to extend the field $\phi(x)$ to $\phi(x, \tau)$, which evolves according to a Langevin equation,
\begin{equation}
    \frac{\partial \phi(x, \tau)}{\partial \tau} = -\frac{\delta S[\phi]}{\delta \phi(x, \tau)} + \sqrt{2}\eta(x, \tau),
\end{equation}
where $\eta(x, \tau)$ is a Gaussian white noise term. This reformulation allows for the evolution towards a stationary distribution which corresponds to the (Euclidean) path integral of the quantum field theory.

Diffusion models consist of a forward diffusion process and a reverse denoising process. The forward process follows the stochastic differential equation 
$ \frac{\partial \phi}{\partial\xi} = f(\phi, \xi) + g(\xi) \eta(\xi), $
where $0\leq\xi\leq T$ is the forward time, $f(\phi, \xi)$ is the drift term, $g(\xi)$ is the diffusion term, and $\eta(\xi)$ is noise. The reverse process~\cite{anderson:1982reversetime} is described by 
$ \frac{\partial\phi}{\partial t} = \left[ f(\phi, t) - g^2(t)\nabla_{\phi}\log p_t(\phi) \right] + g(t) \eta(t), $
where $t$ flows backward from $T$ to 0, and $p_t(\phi)$ is the distribution of $\phi(x)$ at time $t$. Although Gaussian noise in the forward process simplifies the reverse process, modeling the $t$-dependent drift term requires a deep neural network, which approximates it as 
$ s_{\hat{\theta}}(\phi,t) \simeq \left[ f(\phi, t) - g^2(t)\nabla_{\phi}\log p_t(\phi) \right]. $
This allows sample generation via the reverse process 
\begin{equation}
  \frac{\partial\phi(x,t)}{\partial t} =  s_{\hat{\theta}}(\phi(x,t),t) + g(t)\eta(x,t), \label{eq:sde}
\end{equation}
starting from a normal prior for $\phi(x)$, and with $\eta(x,t)$ again Gaussian white noise.

In statistical systems and field theories, configurations are sampled from the physical distribution $p(\phi) = \exp[-S(\phi;\beta)]/Z$, where $\beta$ is a parameter. If the action is separable, $S(\phi;\beta) = \beta \tilde S(\phi)$, one can carry out simulations at a reference value $\beta_0$ and obtain results at different values of $\beta$ using Ferrenberg-Swendsen or histogram reweighting \cite{PhysRevLett.61.2635}, provided there is sufficient overlap between the simulated and the target ensembles.

In diffusion models, with the variance-expanding scheme~\cite{song2021scorebased}, the network represented score function $s_\theta$ is trained to learn the drift term $ - g^2(t)\nabla_{\phi}\log p_t(\phi)$ in the reverse process. In Langevin simulations of fields, the drift term is determined by the action, $-{\delta S[\phi]}/{\delta \phi(x, \tau)}$. If the action is separable, as in pure gauge theories, the action and its derivative are proportional to the gauge coupling $\beta$. This suggests that the well-learned score function at a reference coupling $\beta_0$ can be used to replace the drift term in Langevin simulations as
\begin{equation}
\label{eq:reweight}
 \frac{\partial\phi(x,t)}{\partial t} =  \frac{\beta}{\beta_0} s_{\hat{\theta}}(\phi(x,t),t;\beta_0) + g(t)\bar{\eta}(x,t), 
\end{equation}
for sample generation at arbitrary $\beta$. Note that this extrapolation also avoids the need to estimate the $\beta$-dependent normalization factor, since $\nabla_{\phi}\log p(\phi) \equiv -\nabla_{\phi} (S[\phi] + \ln{Z(\beta)}) = -\nabla_{\phi}S[\phi] $. We will refer to this formulation as a physics-conditioned diffusion model.

\section{Gauge fields}
\label{sec:gauge}

\subsection{Abelian gauge field theory}

Abelian or $U(1)$ gauge theories in two Euclidean dimensions provide simplified models for gauge theories~\cite{Smit:2002ug,Kanwar:2020xzo,Crean:2024nro}, while retaining some of the essential features of nonabelian theories in four dimensions. The (continuum) action is $S = \frac{1}{4} \int d^2x \, F_{\mu\nu}F^{\mu\nu}$, 
with the field strength tensor $F_{\mu\nu} = \partial_\mu A_\nu - \partial_\nu A_\mu$ and $U(1)$ vector potentials $A_\mu$ (with $\mu = 0, 1$).

On an $L \times L$ square lattice with periodic boundary conditions, gauge fields are represented by link variables $U_{x,\mu} = e^{i \theta_{x,\mu}}$, where $\theta_{x,\mu}$ is the gauge angle for the link from site $x$ to $x + \hat{\mu}$. The field strength is related to the plaquette variable $U_{\Box} \equiv U_{x,\mu} U_{x+\hat{\mu},\nu} U_{x+\hat{\nu},\mu}^\dagger U_{x,\nu}^\dagger = e^{i F_{01}(x)}$, and the lattice action is
\begin{equation}
    S = -\beta \sum_{\Box} \text{Re}(U_{\Box}).
\end{equation}
Here $\beta$ is the inverse coupling which controls the gauge interaction strength and ensures consistency with the continuum limit (at large $\beta$) in simulations.

Physical properties of interest are estimated via the expectation value of observables $\hat{O}$ using the Euclidean path integral, $\langle \hat O \rangle = \int \mathcal{D}U\, O(U) e^{-S(U)}/Z,$ where $\int \mathcal{D}U$ denotes integration over the Haar measure for each link and $Z = \int \mathcal{D}U\, e^{-S(U)}$. The integration corresponds to the statistical expectation value over all configurations in the distribution. As observables, we consider here Wilson loops $W(C)$, the topological charge $Q$, and the topological susceptibility $\chi_Q$.

The Wilson loop $W(C)$ is a gauge-invariant observable defined as the (trace of the path-ordered) product of link variables around a closed contour $C$, $W(C) = \text{Tr} \left( \mathcal{P} \exp \left( i \oint_C A_\mu dx^\mu \right) \right)$. 
For abelian theories, the trace and path ordering are not required. 
On the lattice, this becomes
\begin{equation}
    W(C) = \text{Tr} \left( \prod_{(x,\mu) \in C} U_{x,\mu} \right). 
\end{equation}
For a rectangular loop of size $\mathit{l}\times \mathit{l}$, 
$W_{\mathit{l}\times \mathit{l}}$ measures the flux through the region enclosed by $C$. In two dimensions, the topological charge $Q$ characterizes the winding number of the gauge field configuration and is on the lattice given by
\begin{equation}
    Q = \frac{1}{2\pi} \sum_{x} F_{01}(x),
\end{equation}
where the phase $F_{01}$ of the plaquette is chosen within the principal interval, i.e.\ $F_{01}(x) \in (-\pi, \pi]$. The topological charge is significant in gauge theories as it relates to flux quantization and instantons. The topological susceptibility is given by $\chi_Q = \langle Q^2/V \rangle$, where $V = L^2$ is the two-dimensional volume.

\subsection{Numerical tests}

To illustrate our approach, we examine a fixed lattice size of $L = 16$ and train the DM at $\beta = 1$, using 30,720 configurations generated with hybrid Monte Carlo (HMC). The gauge angles $\theta_{x,\mu}$ are the chosen degrees of freedom. We apply a series of gauge transformations to the original configurations, treating the transformed configurations as new configurations for training. By performing four gauge transformations on the original set, we increase the total number of configurations used for training to 153,600.

With regard to the set-up of DM, we represent the score function with a U-Net architecture following our previous work~\cite{Wang:2023exq}, with an encoder-decoder structure and symmetric skip connections. The default shape of the input is set as $(\cdot,2,16,16)$, where the first dimension is the batch size, the second indicates the channel, and the final two represent dimensionalities of the field configuration. The skip connections between corresponding encoder and decoder layers help preserve spatial information lost during downsampling. To account for the periodic boundary conditions in Lattice QCD configurations, periodic padding is applied at each layer of the network and time and spatial directions are viewed as two distinct channels of the network. We subsequently test the well-trained DM to generate configurations at different $\beta$ values with physics-conditions.

%%%%%%%%%%%%%%%%%%%%%%%%%%%%%%%%%%%%%%%%%%%%%%%%%%%%%%%%%%%%%%%
\begin{figure}[!htbp]
\begin{center}
    \includegraphics[width=0.45\textwidth]{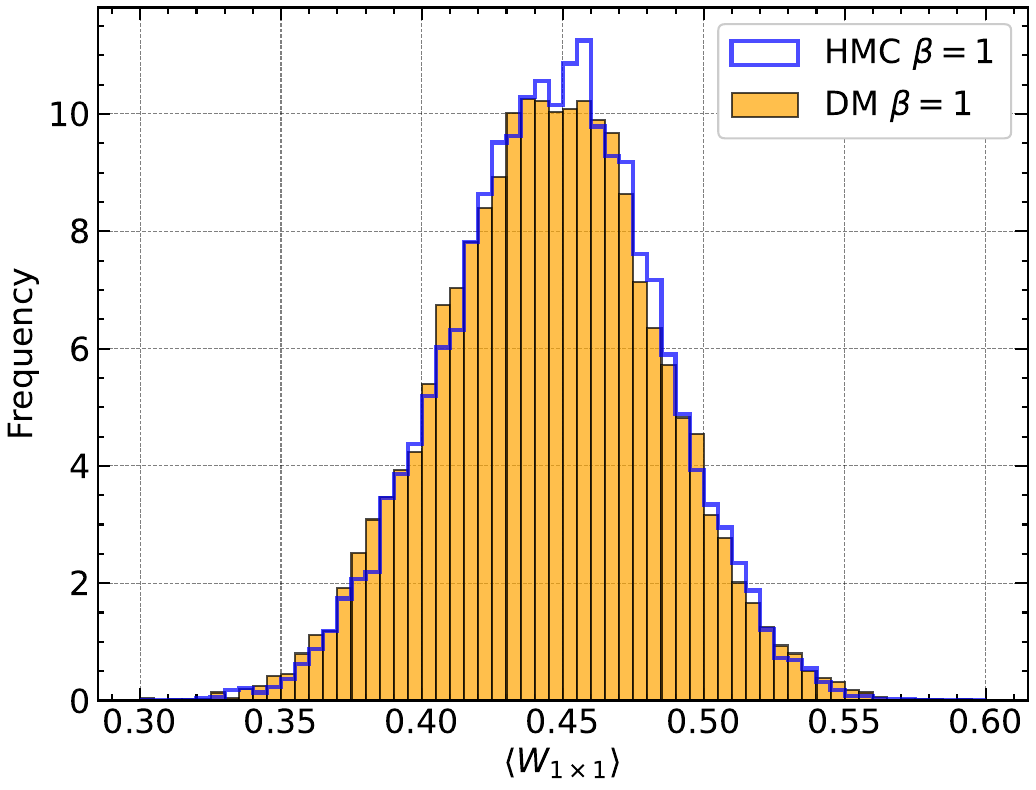}
    \includegraphics[width=0.46\textwidth]{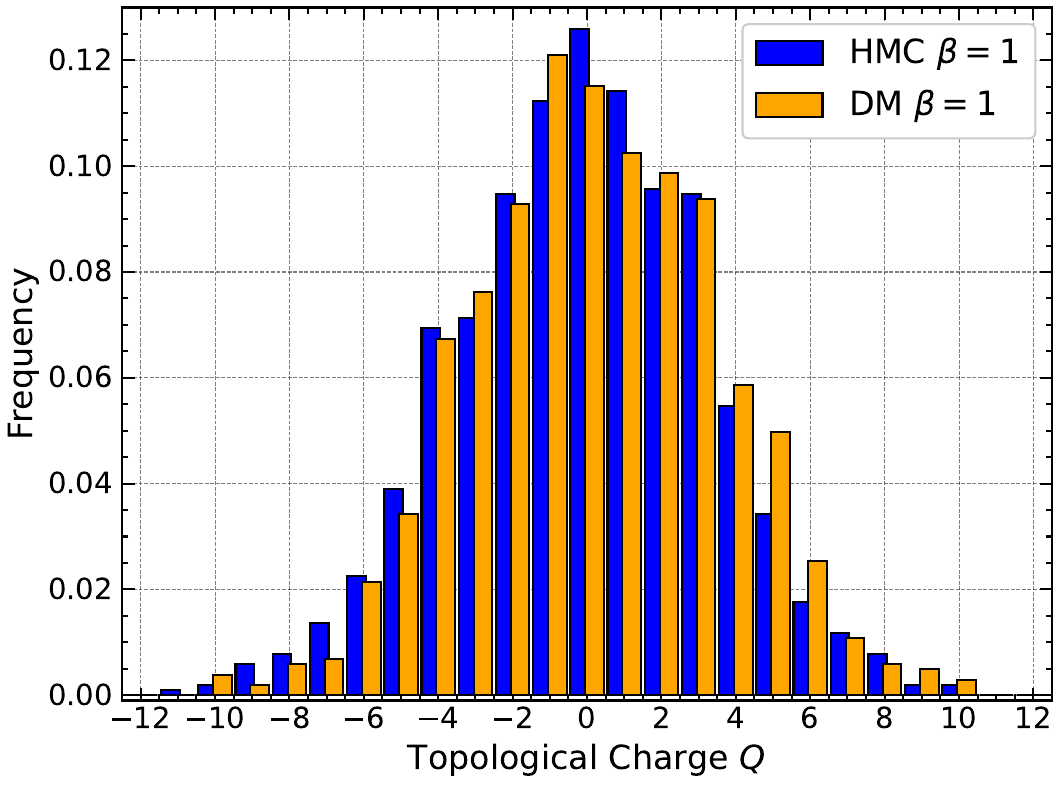}
\end{center}
\caption{Comparison of distributions for the Wilson loop (left)  and the topological charge (right) at $\beta = 1$ from the test data-set (HMC) and from the trained DM. The number of independent configurations is 1,024 in both cases.
}
    \label{fig:beta-1}
\end{figure}
%%%%%%%%%%%%%%%%%%%%%%%%%%%%%%%%%%%%%%%%%%%%%%%%%%%%%%%%%%%%%%%

The performance of the training procedure at $\beta = 1$ is presented in Figure~\ref{fig:beta-1}. Using HMC we generated an additional 1,024 configurations as our test ensemble and compared these with 1,024 configurations generated by the trained DM. The distributions of the Wilson loop (left) are consistent. The distributions of the topological charge (right) are consistent as well and, moreover, we note that no issues related to topological freezing are observed in either the HMC or DM results, as evidenced by the wide distribution of topological charges.

%%%%%%%%%%%%%%%%%%%%%%%%%%%%%%%%%%%%%%%%%%%%%%%%%%%%%%%%%%%%%%%
\begin{figure}[!htbp]
\begin{center}
    \includegraphics[width=0.45\textwidth]{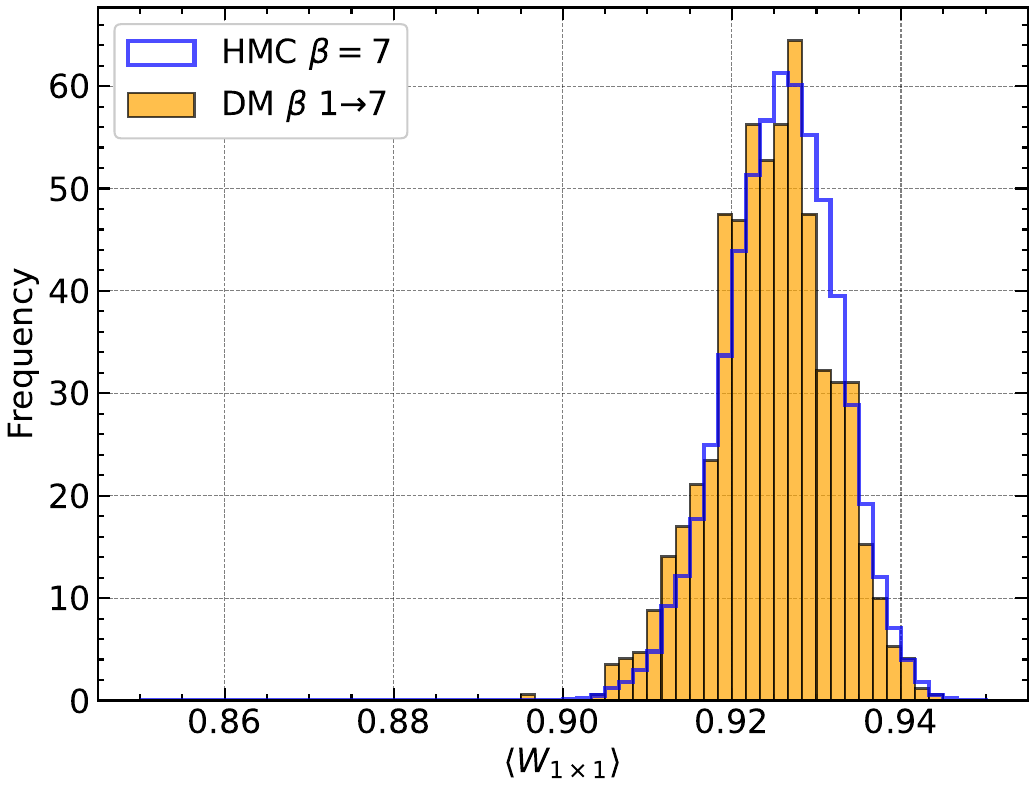}
    \includegraphics[width=0.45\textwidth]{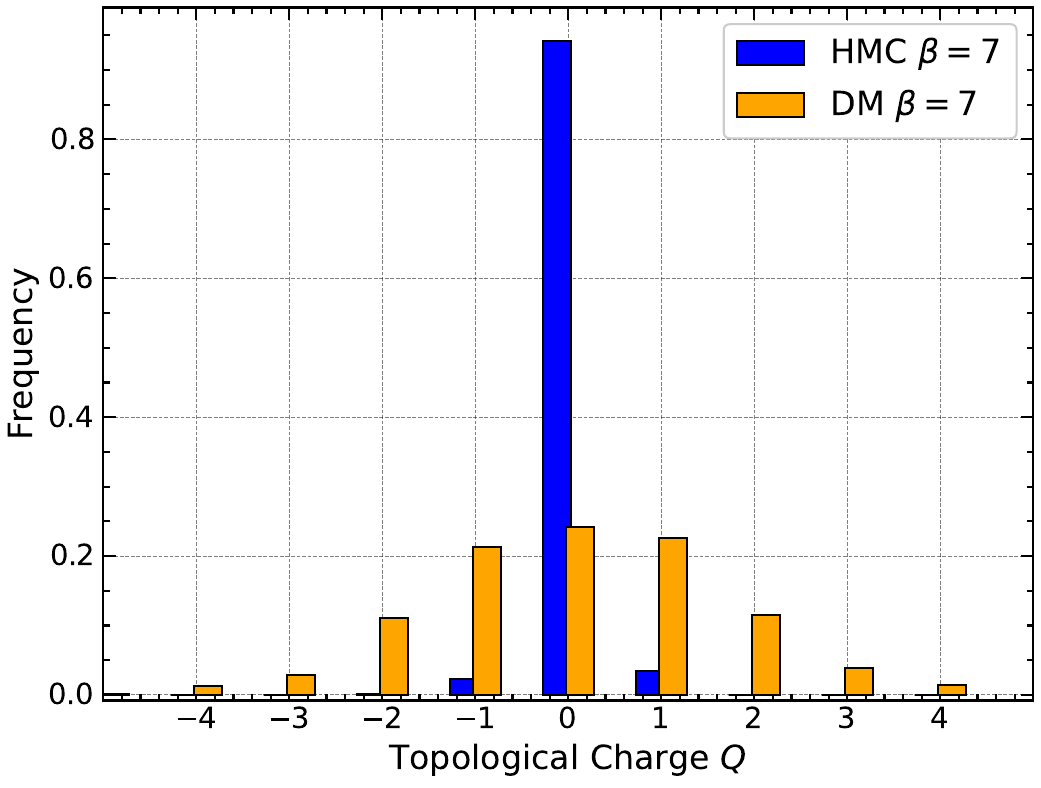}
\end{center}
\caption{
    Comparison of distributions for the Wilson loop (left)  and the topological charge (right) at $\beta = 7$ from the test data-set (HMC) and from the DM trained at $\beta=1$ but conditioned at $\beta=7$. The number of independent configurations is 1,024 in both cases.
}
\label{fig:beta-7}
\end{figure}
%%%%%%%%%%%%%%%%%%%%%%%%%%%%%%%%%%%%%%%%%%%%%%%%%%%%%%%%%%%%%%%

Next we use the DM trained at $\beta=1$ to generate 1,024 configurations at $\beta=7$, using Eq.~(\ref{eq:reweight}), and also use HMC to generate an equal number of configurations directly at $\beta=7$. The results are displayed in Figure~\ref{fig:beta-7}. 
We observe that the Wilson loop distributions (left) remain consistent between the two appraoches. For the topological charge (right) on the other hand, we observe that HMC suffers from topological freezing, only sampling values of $Q$ at or around $0$ . In contrast, the physics-conditioned DM is able to explore a wider range of topological sectors, yielding a larger topological susceptibility. We are currently comparing the numerically computed distribution with the analytical prediction, which is possible in this simple theory \cite{Kanwar:2021wzm}.

%%%%%%%%%%%%%%%%%%%%%%%%%%%%%%%%%%%%%%%%%%%%%%%%%%%%%%%%%%%%%%%
\begin{figure}[htbp!]
    \centering
    \includegraphics[width=1.\linewidth]{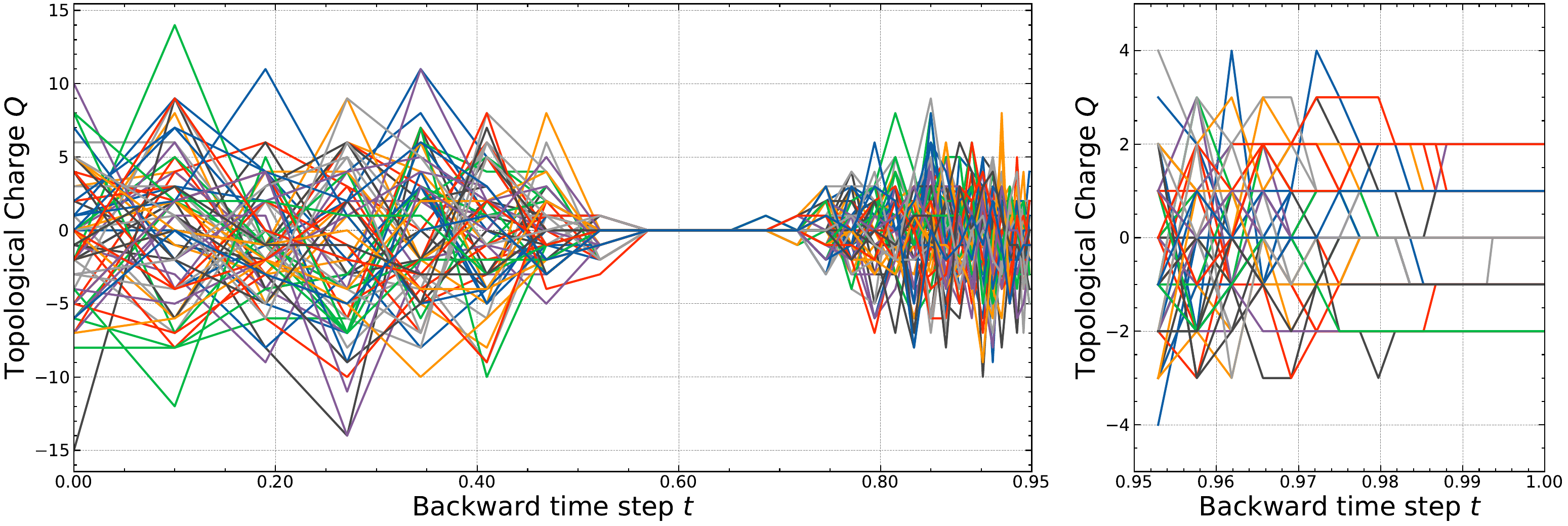}
    \caption{Trajectories of topological charge $Q$ for 64 different initial configurations, generated in the backward process at $\beta = 7$, using the diffusion model trained at $\beta = 1$. The left panel gives the evolution from 0 to 0.95. The right panel shows the rest time steps with a zoomed plot.}
    \label{fig:traj}
\end{figure}
%%%%%%%%%%%%%%%%%%%%%%%%%%%%%%%%%%%%%%%%%%%%%%%%%%%%%%%%%%%%%%%

To see how the DM samples the topological sectors, we show in Figure~\ref{fig:traj} trajectories of topological charge $Q$ for 64 different initial configurations, generated in the backward process at $\beta = 7$, using the diffusion model trained at $\beta = 1$.
We observe that in the first part of the backward process, up to $t~\sim 0.5$, the topological charge is fluctuating substantially, before all 64 trajectories converge to $Q=0$. The physically relevant part of the evolution occurs towards the end of the backward process, where the different trajectories explore the topological sectors again and finally settle down to yield a distribution similar to the one shown in Figure \ref{fig:beta-7} (right).

Besides the above computations, the detailed implementation of HMC, DM, and Langevin simulations, along with a thorough analysis of their performance and associated uncertainties, will be presented in a coming comprehensive study. This future work will provide an in-depth discussion of the methodologies employed, as well as a rigorous assessment of the accuracy and reliability of the simulations.

\section{Summary}

Diffusion models are a promising tool to generate configurations in lattice field theory. Here we have presented diffusion models for lattice gauge field simulations, and introduced a new approach to extrapolate in the gauge coupling, by adjusting the learned score by a simple multiplication, as motivated by the connection to stochastic quantization.
We have presented an implementation for a $U(1)$ theory in two dimensions and shown that the issue of topological freezing is alleviated when compared to the traditional HMC algorithm, by employing a diffusion model trained in a region of parameter space where topological freezing is absent.

\noindent
{\bf Acknowledgements}  -- We thank Drs.\ Akinori Tanaka, Tetsuo Hatsuda, Andreas Ipp, Yin Jiang, Gurtej Kanwar, Fernando Romero-L\'opez, and Shuzhe Shi for helpful discussions.
% Institutions
We thank the DEEP-IN working group at RIKEN-iTHEMS for support in the preparation of this paper.
% Funds
GA is supported by STFC Consolidated Grant ST/T000813/1. LW also thanks to the National Natural Science Foundation (NSFC) of China under Grant No.\ 12147101 for supporting his visit to Shanghai Research Center for Theoretical Nuclear Physics, NSFC and Fudan University. KZ is supported by the CUHK-Shenzhen university development fund under grant No.\ UDF01003041 and UDF03003041, and Shenzhen Peacock fund under No.\ 2023TC0179. WW and QZ are supported in part by Natural Science Foundation of China under grants No.\ 12125503 and 12335003, and SJTU Kunpeng \& Ascend Center of Excellence.

\noindent
{\bf Open Access Statement} -- For the purpose of open access, the authors have applied a Creative Commons Attribution (CC BY) licence to any Author Accepted Manuscript version arising.

\newpage

\bibliographystyle{JHEP.bst}
\bibliography{refs}

\providecommand{\href}[2]{#2}\begingroup\raggedright\begin{thebibliography}{10}

\bibitem{Ding:2015ona}
H.-T.~Ding, F.~Karsch and S.~Mukherjee, \emph{{Thermodynamics of
  strong-interaction matter from Lattice QCD}},
  \href{https://doi.org/10.1142/S0218301315300076}{\emph{Int. J. Mod. Phys. E}
  {\bfseries 24} (2015) 1530007}
  [\href{https://arxiv.org/abs/1504.05274}{{\ttfamily 1504.05274}}].

\bibitem{Gross:2022hyw}
F.~Gross et~al., \emph{{50 Years of Quantum Chromodynamics}},
  \href{https://doi.org/10.1140/epjc/s10052-023-11949-2}{\emph{Eur. Phys. J. C}
  {\bfseries 83} (2023) 1125}
  [\href{https://arxiv.org/abs/2212.11107}{{\ttfamily 2212.11107}}].

\bibitem{Aarts:2023vsf}
G.~Aarts et~al., \emph{{Phase Transitions in Particle Physics}: {Results and
  Perspectives from Lattice Quantum Chromo-Dynamics}},
  \href{https://doi.org/10.1016/j.ppnp.2023.104070}{\emph{Prog. Part. Nucl.
  Phys.} {\bfseries 133} (2023) 104070}
  [\href{https://arxiv.org/abs/2301.04382}{{\ttfamily 2301.04382}}].

\bibitem{Alles:1996vn}
B.~Alles, G.~Boyd, M.~D'Elia, A.~Di~Giacomo and E.~Vicari, \emph{{Hybrid Monte
  Carlo and topological modes of full QCD}},
  \href{https://doi.org/10.1016/S0370-2693(96)01247-6}{\emph{Phys. Lett. B}
  {\bfseries 389} (1996) 107}
  [\href{https://arxiv.org/abs/hep-lat/9607049}{{\ttfamily hep-lat/9607049}}].

\bibitem{DelDebbio:2004xh}
L.~Del~Debbio, G.M.~Manca and E.~Vicari, \emph{{Critical slowing down of
  topological modes}},
  \href{https://doi.org/10.1016/j.physletb.2004.05.038}{\emph{Phys. Lett. B}
  {\bfseries 594} (2004) 315}
  [\href{https://arxiv.org/abs/hep-lat/0403001}{{\ttfamily hep-lat/0403001}}].

\bibitem{Albandea:2021lvl}
D.~Albandea, P.~Hern\'andez, A.~Ramos and F.~Romero-L\'opez, \emph{{Topological
  sampling through windings}},
  \href{https://doi.org/10.1140/epjc/s10052-021-09677-6}{\emph{Eur. Phys. J. C}
  {\bfseries 81} (2021) 873}
  [\href{https://arxiv.org/abs/2106.14234}{{\ttfamily 2106.14234}}].

\bibitem{Albandea:2021kwe}
D.~Albandea, P.~Hern\'andez, A.~Ramos and F.~Romero-L\'opez, \emph{{Improved
  topological sampling for lattice gauge theories}},
  \href{https://doi.org/10.22323/1.396.0183}{\emph{PoS} {\bfseries LATTICE2021}
  (2022) 183} [\href{https://arxiv.org/abs/2111.05745}{{\ttfamily
  2111.05745}}].

\bibitem{Cossu:2021bgn}
G.~Cossu, D.~Lancastera, B.~Lucini, R.~Pellegrini and A.~Rago, \emph{{Ergodic
  sampling of the topological charge using the density of states}},
  \href{https://doi.org/10.1140/epjc/s10052-021-09161-1}{\emph{Eur. Phys. J. C}
  {\bfseries 81} (2021) 375}
  [\href{https://arxiv.org/abs/2102.03630}{{\ttfamily 2102.03630}}].

\bibitem{Parisi:1980ys}
G.~Parisi and Y.-s.~Wu, \emph{{Perturbation Theory Without Gauge Fixing}},
  {\emph{Sci. Sin.} {\bfseries 24} (1981) 483}.

\bibitem{Damgaard:1987rr}
P.H.~Damgaard and H.~H{\"u}ffel, \emph{Stochastic quantization},
  \href{https://doi.org/10.1016/0370-1573(87)90144-X}{\emph{Phys. Rept.}
  {\bfseries 152} (1987) 227}.

\bibitem{Namiki:1993fd}
M.~Namiki, \emph{{Basic ideas of stochastic quantization}},
  \href{https://doi.org/10.1143/PTPS.111.1}{\emph{Prog. Theor. Phys. Suppl.}
  {\bfseries 111} (1993) 1}.

\bibitem{Cranmer:2023xbe}
K.~Cranmer, G.~Kanwar, S.~Racani\`ere, D.J.~Rezende and P.E.~Shanahan,
  \emph{{Advances in machine-learning-based sampling motivated by lattice
  quantum chromodynamics}},
  \href{https://doi.org/10.1038/s42254-023-00616-w}{\emph{Nature Rev. Phys.}
  {\bfseries 5} (2023) 526} [\href{https://arxiv.org/abs/2309.01156}{{\ttfamily
  2309.01156}}].

\bibitem{Kanwar:2020xzo}
G.~Kanwar, M.S.~Albergo, D.~Boyda, K.~Cranmer, D.C.~Hackett, S.~Racani\`ere
  et~al., \emph{{Equivariant flow-based sampling for lattice gauge theory}},
  \href{https://doi.org/10.1103/PhysRevLett.125.121601}{\emph{Phys. Rev. Lett.}
  {\bfseries 125} (2020) 121601}
  [\href{https://arxiv.org/abs/2003.06413}{{\ttfamily 2003.06413}}].

\bibitem{Boyda:2020hsi}
D.~Boyda, G.~Kanwar, S.~Racani\`ere, D.J.~Rezende, M.S.~Albergo, K.~Cranmer
  et~al., \emph{{Sampling using $SU(N)$ gauge equivariant flows}},
  \href{https://doi.org/10.1103/PhysRevD.103.074504}{\emph{Phys. Rev. D}
  {\bfseries 103} (2021) 074504}
  [\href{https://arxiv.org/abs/2008.05456}{{\ttfamily 2008.05456}}].

\bibitem{Kanwar:2021wzm}
G.~Kanwar, \emph{{Machine Learning and Variational Algorithms for Lattice Field
  Theory}}, Ph.D. thesis, MIT, 2021.
\newblock \href{https://arxiv.org/abs/2106.01975}{{\ttfamily 2106.01975}}.

\bibitem{Albergo:2021vyo}
M.S.~Albergo, D.~Boyda, D.C.~Hackett, G.~Kanwar, K.~Cranmer, S.~Racani\`ere
  et~al., \emph{{Introduction to Normalizing Flows for Lattice Field Theory}},
  \href{https://arxiv.org/abs/2101.08176}{{\ttfamily 2101.08176}}.

\bibitem{Kanwar:2024ujc}
G.~Kanwar, \emph{{Flow-based sampling for lattice field theories}},  in
  \emph{{40th International Symposium on Lattice Field Theory}}, 1, 2024
  [\href{https://arxiv.org/abs/2401.01297}{{\ttfamily 2401.01297}}].

\bibitem{chen:2018}
R.T.~Chen, Y.~Rubanova, J.~Bettencourt and D.K.~Duvenaud, \emph{Neural ordinary
  differential equations}, {\emph{Advances in neural information processing
  systems} {\bfseries 31} (2018) }
  [\href{https://arxiv.org/abs/1806.07366}{{\ttfamily 1806.07366}}].

\bibitem{deHaan:2021erb}
P.~de~Haan, C.~Rainone, M.C.N.~Cheng and R.~Bondesan, \emph{{Scaling Up Machine
  Learning For Quantum Field Theory with Equivariant Continuous Flows}},
  \href{https://arxiv.org/abs/2110.02673}{{\ttfamily 2110.02673}}.

\bibitem{Gerdes:2022eve}
M.~Gerdes, P.~de~Haan, C.~Rainone, R.~Bondesan and M.C.N.~Cheng,
  \emph{{Learning Lattice Quantum Field Theories with Equivariant Continuous
  Flows}},  \href{https://arxiv.org/abs/2207.00283}{{\ttfamily 2207.00283}}.

\bibitem{Caselle:2023mvh}
M.~Caselle, E.~Cellini and A.~Nada, \emph{{Sampling the lattice Nambu-Goto
  string using Continuous Normalizing Flows}}, {\emph{Journal of High Energy
  Physics} {\bfseries 02} (2024) 048}
  [\href{https://arxiv.org/abs/2307.01107}{{\ttfamily 2307.01107}}].

\bibitem{wu2020stochastic}
H.~Wu, J.~K{\"o}hler and F.~No{\'e}, \emph{Stochastic normalizing flows},
  {\emph{Advances in Neural Information Processing Systems} {\bfseries 33}
  (2020) 5933} [\href{https://arxiv.org/abs/2002.06707}{{\ttfamily
  2002.06707}}].

\bibitem{Caselle:2022acb}
M.~Caselle, E.~Cellini, A.~Nada and M.~Panero, \emph{Stochastic normalizing
  flows as non-equilibrium transformations},
  \href{https://doi.org/10.1007/JHEP07(2022)015}{\emph{JHEP} {\bfseries 07}
  (2022) 015} [\href{https://arxiv.org/abs/2201.08862}{{\ttfamily
  2201.08862}}].

\bibitem{yang:2022diffusion}
L.~Yang, Z.~Zhang, S.~Hong, R.~Xu, Y.~Zhao, Y.~Shao et~al., \emph{Diffusion
  {{Models}}: {{A Comprehensive Survey}} of {{Methods}} and {{Applications}}},
  \href{https://arxiv.org/abs/2209.00796}{{\ttfamily 2209.00796}}.

\bibitem{croitoru2023diffusion}
F.-A.~Croitoru, V.~Hondru, R.T.~Ionescu and M.~Shah, \emph{Diffusion models in
  vision: A survey}, {\emph{IEEE Transactions on Pattern Analysis and Machine
  Intelligence} {\bfseries 45} (2023) 10850}.

\bibitem{Mikuni:2022xry}
V.~Mikuni and B.~Nachman, \emph{{Score-based generative models for calorimeter
  shower simulation}},
  \href{https://doi.org/10.1103/PhysRevD.106.092009}{\emph{Phys. Rev. D}
  {\bfseries 106} (2022) 092009}
  [\href{https://arxiv.org/abs/2206.11898}{{\ttfamily 2206.11898}}].

\bibitem{Amram:2023onf}
O.~Amram and K.~Pedro, \emph{{Denoising diffusion models with geometry
  adaptation for high fidelity calorimeter simulation}},
  \href{https://doi.org/10.1103/PhysRevD.108.072014}{\emph{Phys. Rev. D}
  {\bfseries 108} (2023) 072014}
  [\href{https://arxiv.org/abs/2308.03876}{{\ttfamily 2308.03876}}].

\bibitem{Mikuni:2023dvk}
V.~Mikuni, B.~Nachman and M.~Pettee, \emph{{Fast point cloud generation with
  diffusion models in high energy physics}},
  \href{https://doi.org/10.1103/PhysRevD.108.036025}{\emph{Phys. Rev. D}
  {\bfseries 108} (2023) 036025}
  [\href{https://arxiv.org/abs/2304.01266}{{\ttfamily 2304.01266}}].

\bibitem{Devlin:2023jzp}
P.~Devlin, J.-W.~Qiu, F.~Ringer and N.~Sato, \emph{{Diffusion model approach to
  simulating electron-proton scattering events}},
  \href{https://doi.org/10.1103/PhysRevD.110.016030}{\emph{Phys. Rev. D}
  {\bfseries 110} (2024) 016030}
  [\href{https://arxiv.org/abs/2310.16308}{{\ttfamily 2310.16308}}].

\bibitem{Wang:2023exq}
L.~Wang, G.~Aarts and K.~Zhou, \emph{{Diffusion models as stochastic
  quantization in lattice field theory}},
  \href{https://doi.org/10.1007/JHEP05(2024)060}{\emph{JHEP} {\bfseries 05}
  (2024) 060} [\href{https://arxiv.org/abs/2309.17082}{{\ttfamily
  2309.17082}}].

\bibitem{Wang:2023sry}
L.~Wang, G.~Aarts and K.~Zhou, \emph{{Generative Diffusion Models for Lattice
  Field Theory}},  in \emph{{37th Conference on Neural Information Processing
  Systems}}, 11, 2023 [\href{https://arxiv.org/abs/2311.03578}{{\ttfamily
  2311.03578}}].

\bibitem{Hirono:2024zyg}
Y.~Hirono, A.~Tanaka and K.~Fukushima, \emph{{Understanding Diffusion Models by
  Feynman's Path Integral}},
  \href{https://arxiv.org/abs/2403.11262}{{\ttfamily 2403.11262}}.

\bibitem{anderson:1982reversetime}
B.D.O.~Anderson, \emph{Reverse-time diffusion equation models},
  \href{https://doi.org/10.1016/0304-4149(82)90051-5}{\emph{Stochastic
  Processes and their Applications} {\bfseries 12} (1982) 313}.

\bibitem{PhysRevLett.61.2635}
A.M.~Ferrenberg and R.H.~Swendsen, \emph{{New Monte Carlo technique for
  studying phase transitions}},
  \href{https://doi.org/10.1103/PhysRevLett.61.2635}{\emph{Phys. Rev. Lett.}
  {\bfseries 61} (1988) 2635}.

\bibitem{song2021scorebased}
Y.~Song, J.~{Sohl-Dickstein}, D.P.~Kingma, A.~Kumar, S.~Ermon and B.~Poole,
  \emph{Score-based generative modeling through stochastic differential
  equations},  in \emph{Int. {{Conf}}. {{Learn}}. {{Represent}}.}, 2021.

\bibitem{Smit:2002ug}
J.~Smit, \emph{{Introduction to quantum fields on a lattice: A robust mate}},
  vol.~15, Cambridge University Press (1, 2011).

\bibitem{Crean:2024nro}
X.~Crean, J.~Giansiracusa and B.~Lucini, \emph{{Topological Data Analysis of
  Monopole Current Networks in $U(1)$ Lattice Gauge Theory}},
  \href{https://arxiv.org/abs/2403.07739}{{\ttfamily 2403.07739}}.

\end{thebibliography}\endgroup

\end{document}